\documentclass[prl,twocolumn,amsmath,amssymb,superscriptaddress]{revtex4}

\usepackage{graphicx}
\usepackage{hyperref}

\newcommand{\la}{\label}
\newcommand{\be}{\begin{equation}}
\newcommand{\ee}{\end{equation}}
\newcommand{\bea}{\begin{eqnarray}}
\newcommand{\eea}{\end{eqnarray}}

\def\Tr{\mbox{Tr}\,}
\def\tr{\mbox{tr}\,}

\begin{document}

\title{Allowed charge transfers between coherent conductors driven by a 
time-dependent scatterer.}
\author{A.~G.~Abanov}
\affiliation{Department of Physics and Astronomy,
Stony Brook University,  Stony Brook, NY 11794-3800.}
\affiliation{Institute of Theoretical Physics,
Ecole Polytechnique F\'ed\'erale de Lausanne, CH-1015 Lausanne,
Switzerland}
\author{D.~A.~Ivanov}
\affiliation{Institute of Theoretical Physics,
Ecole Polytechnique F\'ed\'erale de Lausanne, CH-1015 Lausanne,
Switzerland}

\begin{abstract}

We derive constraints on the statistics of the charge transfer between
two conductors in the model of arbitrary time-dependent instant scattering
of non-interacting fermions at zero temperature. The constraints are 
formulated in terms of analytic properties of the generating function:
its zeroes must lie on the negative real axis. This result generalizes
existing studies for scattering by a time-independent scatterer
under time-dependent bias voltage.

\end{abstract}


\maketitle

\paragraph{1. Introduction.}
The concept of full counting statistics (FCS) for charge transfer in
coherent conductors has been introduced in Ref.~\onlinecite{1993-LevitovLesovik}.
While the average current through small junctions may usually be understood
in classical terms, its fluctuations reflect the quantum nature of electrons;
even in the model of non-interacting electrons, the fluctuations of the current
are non-classical due to the electronic Fermi statistics \cite{2000-BlanterButtiker}. 
Generally, one studies
the statistical distribution of the charge transfer between several conductors
coupled by a time-dependent scattering matrix. In the simplest case (considered
in the present paper), there are only two leads, and therefore the charge transfer
may be statistically described by the probabilities $p_q$ to transfer exactly $q$
electrons from the left to the right lead (here $q$ is an integer number, either
positive or negative). The leads are assumed to provide a thermal source of incoming
electrons with the Fermi occupation number $n_F(E)$. 
In the context of coherent quantum manipulation, one of the most 
interesting setups is that of an ``adiabatic pumping'' where the scattering 
matrix varies slowly compared to the characteristic scattering time \cite{1998-Brouwer}.
Neglecting the scattering time amounts to considering the scattering matrix
to be independent of the energy $E$, i.e., local in time \cite{1997-IvanovLeeLevitov}. 
In this approximation, in the model of non-interacting electrons, a general
expression (the so-called ``determinant formula'')
for the charge-transfer statistics $\{ p_q \}$ in terms of the time evolution
of the single-particle scattering matrix $S(t)$ has been 
derived \cite{1993-LevitovLesovik,1996-LevitovLeeLesovik}.

Remarkably, a recent progress has been made in understanding the implications of the
old results for the FCS in the particular case of the bias voltage applied to a fixed
scatterer \cite{2007-VanevicNazarovBelzig}. In this ``bias-voltage'' problem,
the scattering matrix $S(t)$ is changed along a one-dimensional trajectory parameterized
by the U(1) phase $\phi(t)=\int V(t) dt$, where $V(t)$ is the applied voltage. Within this
model, non-trivial constraints have been derived for the charge-transfer statistics.
Namely, not every set of probabilities $\{ p_q \}$ is allowed, but only those for which
the zeroes of the generating function
\be
	\chi(\lambda) =\sum_{q=-\infty}^{+\infty}p_q e^{i\lambda q},
\label{gen-function-def}
\ee
obey certain restrictions (see our discussion below).

\paragraph{2. Main result.}

Inspired by this work \cite{2007-VanevicNazarovBelzig}, we derive similar
(weaker) constraints for the more general case of 
adiabatic pumping \cite{1998-Brouwer}: an arbitrary 
time-dependent scattering matrix $S(t)$ with any number of channels. We restrict
our consideration to the case of two conducting leads and to zero temperature.
Under those conditions, we find that the generating function (\ref{gen-function-def})
may take zero values only for a discrete set of $\lambda$ such that $u=e^{i\lambda}$
belongs to the negative real axis $(-\infty,0)$. Furthermore, we find that 
the generating function $\chi(\lambda)$ (and therefore the full charge-transfer
statistics) is uniquely determined by its zeroes $u_k$, up to an overall integer
charge transfer $\chi(u)\mapsto u^N \chi(u)$. These statements constitute the main
finding of our paper.

As a simple illustration of the derived constraint, consider a 
charge-transfer statistics with the only non-zero probabilities
$p_{-1,0,1}=1/3$. While this statistics appears a priori reasonable, it follows
from our results that it can never be produced for any time-dependent scattering matrix
$S(t)$ with any number of channels between two conductors (since such a statistics
would correspond to the pair of complex roots  $u=e^{\pm i 2\pi/3}$).

\paragraph{3. Determinant formula and its regularization.}

The generating function (\ref{gen-function-def}), in the approximation 
of non-interacting electrons and of instant scattering, 
is given by the determinant 
formula \cite{1993-LevitovLesovik,1993-IvanovLevitov,1996-LevitovLeeLesovik,1997-IvanovLeeLevitov},
\be
	\chi(\lambda) =
	 \det\left(1+n_{F}(S^{\dagger}e^{i\lambda P_3}Se^{-i\lambda P_3}-1)\right)\, ,
 \la{LevF1}
\ee
where $S$ is the time-dependent single-particle scattering matrix. At each moment of time,
$S(t)\in U(2M)$, where $M$ is the number of conducting channels in each of the two conductors.
The operator $P_3=\mbox{diag}(\mathbf{1}_M,\mathbf{0}_M)$ is the projector
counting the charge in the right conductor. The operator $n_F$ is the Fermi 
distribution function in the conductors. 
In this paper, we consider only the case of zero temperature, and so
$n_F(E) = \theta(-E)$ is the step function in the frequency representation
(in the time representation, $n_{F}(t,t')=i/[2\pi(t-t'+i0)]$).
Since the electrons are non-interacting, we treat them as spinless fermions
(the spin may be trivially included as a channel index).

The determinant (\ref{LevF1}) is understood as that of an infinite-dimensional operator
acting both in the time domain and in the $2M$-channel space. In its original form (\ref{LevF1}),
the determinant is ill-defined: while the operator tends to $\mathbf{1}$ at infinite positive energies, it goes to another unitary matrix, $S^{\dagger}e^{i\lambda P_3}Se^{-i\lambda P_3}$, at infinite negative energies.
Therefore the determinant requires a proper regularization \cite{1997-IvanovLeeLevitov}. A consistent
way to regularize the determinant has been proposed in Ref.~\cite{2007-AvronBachmannGrafKlich} 
(their regularization is also equivalent to that used in Ref.~\onlinecite{2003-MuzykantskyAdamov}). 
At zero temperature, the Fermi distribution operator is a projector $n_F^2=n_F$, and 
the regularized determinant may be written in a particularly simple form \cite{2007-AvronBachmannGrafKlich},
\be
	\chi(\lambda) =
	 \det\left\{e^{-i\lambda P_3 n_F}
	 \left[1-P_3+P_3S e^{i\lambda n_F}S^\dagger\right]\right\}.
 \la{LevF2}
\ee

There is a subtle point in calculating the FCS in the situation where charge counting is
performed over a finite time interval: in this case, there are contributions to the noise arising
from the starting and ending of the measurement, even in the absence of 
scattering \cite{1993-LevitovLesovik,1996-LevitovLeeLesovik}. This noise grows
logarithmically with the observation time. To eliminate this contribution from
the switching-on and switching-off, one often considers a {\it periodic} signal $S(t)$
(e.g., by repeating the same signal with a large time period). In this periodic formulation,
at large number of periods $N_p$, the asymptotic behavior of $\chi(\lambda)$ is 
given by \cite{1997-IvanovLeeLevitov}
\be
\chi(\lambda) \sim [\chi_0(\lambda)]^{N_p} \, ,
\ee
where $\chi_0(\lambda)=\exp [\lim_{N_p\to\infty} \frac{1}{N_p} \ln \chi(\lambda)]$ is the
counting statistics {\it per period}. The generating function $\chi(\lambda)$ may be
computed with the same formula (\ref{LevF2}), but with the finite time interval closed
into a loop with the periodic boundary conditions. We further consider this periodic setup
and omit the subscript $0$ by using the notation $\chi(\lambda)$ for the charge-transfer
statistics per period.

To write the operators in (\ref{LevF2}) in the matrix form, one can perform a Fourier
transform in time (introducing discrete quasi-energies). In the quasi-energy basis, the
operators are represented as infinite-dimensional discrete matrices. 
We assume that the dependence $S(t)$ is sufficiently smooth,
so that it induces only short-range transitions between quasi-energies \cite{smooth}.
Then the operator in (\ref{LevF2}) rapidly approaches $\mathbf{1}$ away
from the Fermi level, and the determinant converges for all values of $\lambda$.
Thus we conclude that $\chi(u)$ is a single-valued analytic function of 
$u=e^{i\lambda}$ for any $u\in C \backslash \{ 0 \}$. 

\paragraph{4. Determinant formula in $z$-representation.}
For our discussion, it will be convenient to rewrite (\ref{LevF2}) in a different form. 
We parametrize the $2M\times 2M$ unitary $S$-matrix 
by two complex $2M\times M$ matrices $z$ and $\tilde{z}$ as 
\be
	S^\dagger = (z | \tilde{z}) \, ,
\la{Sz}
\ee
subject to the constraints 
$z^\dagger z=\tilde{z}^\dagger \tilde{z}=\mathbf{1}_M$ and $z^\dagger \tilde{z} =\mathbf{0}_M$. 
These constraints guarantee the unitarity of the $S$-matrix. 

Substituting (\ref{Sz}) into (\ref{LevF2}) and using the explicit form of the charge operator $P_3$,
we immediately obtain
\be
	\chi(\lambda) =  \det\left(e^{-i\lambda n_{F}} 
	z^{\dagger}e^{i\lambda n_{F}}z \right).
\la{zformula}
\ee
Notice that the determinant size in the channel space is $M\times M$ in (\ref{zformula}) versus $2M\times 2M$ in (\ref{LevF2}).

Furthermore, $\chi(\lambda)$ has a symmetry with respect to the right gauge transformation
$z \mapsto z U_{M}(t)$ for any time-dependent unitary $M\times M$ matrix $U_M (t)$. Under
this transformation, $\chi(\lambda)$ gets multiplied by 
$\det\left(U_{M}e^{-i\lambda n_{F}}U_{M}^{\dagger}e^{i\lambda n_{F}}\right)
=\exp [i\lambda\Tr (n_{F}-U_{M}n_{F}U_{M}^{\dagger})]
= \exp\left[i\lambda \int \frac{dt}{2\pi}\,\tr(U_{M}i\partial_{t}U_{M}^{\dagger})\right]$.
Due to the periodicity of $U_{M}(t)$, the integral in the last expression is an integer number,
and it affects only the overall shift of the total pumped charge. 
We conclude, that, up to this integer, the FCS depends only on $z$ modulo $U(M)$ 
gauge transformations. The latter space may be described as $U(2M)/[U(M){\times}U(M)]$
and contains $2M^2$ real parameters [instead of $4M^2$ real parameters in $S\in U(2M)$]. It may be convenient to parameterize it with
$\hat{N}(t)=2zz^{\dagger}-1$, a $2M\times 2M$ traceless Hermitian matrix
obeying the constraint $\hat{N}^2(t)=\mathbf{1}_{2M}$ (this matrix was introduced in
Ref.~\onlinecite{2001-MakhlinMirlin}). Physically, the gauge invariance described above
corresponds to the independence of FCS on the scattering separately in the right outgoing states \cite{2001-MakhlinMirlin}.

In addition to gauge rotations, (\ref{zformula}) is also insensitive 
to global (time-independent) left rotations $z\mapsto U_{2M} z$,
which correspond to global rotations of $\hat{N}(t)$ \cite{2001-Levitov}.

\paragraph{5. Derivation of the main result.}

To derive our main result we rewrite (\ref{zformula}) as 
\be
	\chi(u) = \det\left(1+(u^{-1}-1) n_{F}\right)\left(1+(u-1) n_{F}^z\right),
\la{nFpform}
\ee
where $n_{F}^z=z^{\dagger}n_{F}z$, and
we have used $n_{F}^{2}=n_{F}$.

The first factor in (\ref{nFpform}) does not involve any information about the time-dependent 
scattering. Its only role is to provide a reference point: the position of the Fermi level.  
The dependence on the evolution of $S$-matrix enters through  $n_{F}^z$, which has
a $M{\times}M$ matrix structure and tends to $\mathbf{0}_M$ and to $\mathbf{1}_M$ at
infinite positive and infinite negative energies, respectively.

We can show that (\ref{nFpform}) is fully determined by the spectrum of $n_{F}^z$,
up to an overall constant charge transfer. Indeed, we can rewrite (\ref{nFpform}) as
\bea
	\chi(\lambda) &=& \det\left[\left( e^{-i\lambda n_F} e^{i\lambda n_F^z}\right)
\left(e^{-i\lambda n_F^z} [1+(u-1)n_F^z]\right) \right] \nonumber\\
&=&  \det\left( e^{-i\lambda n_F} e^{i\lambda n_F^z}\right)
\det\left(e^{-i\lambda n_F^z} [1+(u-1)n_F^z]\right) \nonumber\\
&=& e^{i\lambda\Tr(n_F^{z}-n_F)} \det\left(e^{-i\lambda n_F^z} [1+(u-1)n_F^z]\right)\, ,
\la{anomaly-separated}
\eea
where the determinant of the product can be written as the product of the determinants, since
both operators behave properly (tend to $\mathbf{1}_M$) at infinite energies, and the
last transformation is justified, since $n_F^{z}-n_F$ rapidly tends to zero at infinite
energies. The trace in the first factor of the last expression equals the total transferred
charge (generally non-integer) \cite{2007-AvronBachmannGrafKlich},
\be
\Tr(n_F^{z}-n_F)=Q\, .
\label{total-charge}
\ee
The second factor in (\ref{anomaly-separated}) now depends {\it only} on the spectrum of $n_{F}^z$.

For a periodic time dependence of $S(t)$, the operator $n_{F}^z$ may be represented in the
quasi-energy basis as a discrete infinite Hermitian matrix. It tends to $\mathbf{0}$
at large positive energies and to $\mathbf{1}$ at large negative energies.
Its spectrum is discrete and takes real values between 0 and 1, with possible accumulation points at
0 and 1 \cite{discrete-spectrum}. 
If we denote the eigenvalues of $n_{F}^z$ as $n_\alpha^z$, 
Eq.(\ref{anomaly-separated}) implies the following 
expression for the generating function:
\be
\chi(u)=e^{i\lambda Q}\prod_\alpha e^{-i\lambda n_\alpha^z} [1+(u-1)n_\alpha^z]\, .
\label{chi-result1}
\ee
Notice that while this formula involves $\lambda=-i\ln u$, it is
a single-valued function of $u$ on $C\backslash\{0\}$. As explained above, this expression
is valid up to a constant integer charge transfer.

It is obvious from (\ref{nFpform}) that
the spectrum of $n_{F}^z$ is in one-to-one correspondence with the positions of zeroes of $\chi(u)$,
\be
u_\alpha=1-(n_\alpha^z)^{-1}\, .
\label{uz}
\ee
Since the spectrum of $n_{F}^z$ is real and lies between 0 and 1, we conclude that the zeroes
of $\chi(u)$ must all lie on the negative real axis. This argument finishes the demonstration
of the main result of this paper.

\paragraph{6. Example of forbidden charge transfers.}

We can illustrate our result with an example of a charge transfer which
cannot be realized in our scattering system. Consider the following
particular example of FCS:
\be
	\chi(u) = 1-2F +F(u+u^{-1}),
\la{Fex}
\ee
where $0\leq F \leq 1/2$. The corresponding non-vanishing probabilities 
$p_{\pm 1}=F$, $p_{0}=1-2F$ are normalized and non-negative. 
However, the roots of (\ref{Fex}) are real only for $F\leq 1/4$. 
Therefore, the statistics (\ref{Fex}) with
$1/4 < F\leq 1/2$ cannot be produced by any
evolution of scattering matrix at zero temperature. 

\paragraph{7. Allowed charge transfers.}

While the condition of real negative zeroes of $\chi(u)$ is a necessary
condition for an allowed FCS, it is also apparently a sufficient condition.
Indeed, for any finite set of negative real $u_\alpha$,
the statistics
\be
\chi(u)= u^{N_1} \prod_{\alpha=1}^{N_2} \frac{u-u_\alpha}{1-u_\alpha}
\label{chi-result2}
\ee
(which is identical to (\ref{chi-result1}) with $N_1=Q-\sum_\alpha n^z_\alpha$)
can be trivially realized in a system with $N_2+1$ channels by using
single-particle-transfer pulses as described in 
Ref.~\onlinecite{1997-IvanovLeeLevitov} (one channel per each $u_\alpha$ plus
one channel for the overall integer transfer $N_1$). 

Moreover, this FCS may also be arbitrarily closely approximated in a 
{\it single-channel} system by sending individual well-separated pulses
for each of the required zeroes $u_\alpha$. Indeed, in a single-channel system,
the FCS is determined by the time evolution of the vector 
$\Tr \hat{N}\mathbf{\sigma}$ on the two-dimensional sphere,
as shown in Ref.~\onlinecite{2001-MakhlinMirlin}). Motion of this vector along a circular trajectory of a given area (on the sphere) is equivalent to the ``bias-voltage'' problem with a fixed channel transparency determined by the area enclosed by the contour.
This ``bias-voltage'' problem has been considered
in Ref.~\onlinecite{1997-IvanovLeeLevitov} where it has been shown that an
arbitrarily sharp Lorentzian voltage pulse produces the elementary charge transfer
$\chi(u)=(u-u_\alpha)/(1-u_\alpha)$. By superimposing such pulses, well separated in time,
along circles of different areas, we can approximate the required statistics
(\ref{chi-result2}) arbitrarily close.

Furthermore, since we can make this approximation for any finite number
of roots $N_2$, and the possible infinite sets of roots have accumulation
points only at $u_\alpha\to 0 $ and $u_\alpha \to -\infty$, we can also
approximate arbitrarily close any statistics (\ref{chi-result1}) with
an infinite set of roots, provided they converge sufficiently rapidly
to $n^z_\alpha \to 0$ and $n^z_\alpha\to 1$ (for the convergence of
(\ref{chi-result1}) it is sufficient to require that  $n^z_\alpha$
converge not slower than $|\alpha|^{-\eta}$ with $\eta>1$ at their
accumulation points). We leave it as a mathematical problem to determine
the precise requirements on the convergence and the conditions under which
the generating function (\ref{chi-result1}) may be reproduced {\it exactly}
by a suitable evolution of the $S$-matrix.

\paragraph{8. Bias voltage case.}

With this result in mind, we would like to comment on the existing results for
the problem with a {\it restricted} evolution of the scattering matrix, namely
the ``bias-voltage'' case \cite{2007-VanevicNazarovBelzig}. 
In this restricted problem, the transparencies of the
channels are fixed, and the evolution of the $S$-matrix is determined
by the single parameter $\phi(t)=\int V(t) dt$. The time dependence of
$z$ in this setup is given by $z=e^{i\phi(t)P_{3}}z_{0}$, where $z_0$ is a fixed
$2M\times M$ matrix. As follows from the results of Ref.~\onlinecite{2007-VanevicNazarovBelzig},
for this restricted evolution of the scattering matrix, there is an additional
constraint on the positions of $u_\alpha$. Namely, there are two types of
roots $u_\alpha$: ordinary (``typical'' in notation of Ref.~\onlinecite{2007-VanevicNazarovBelzig})
and anomalous. Ordinary roots come in inversion-symmetric pairs ($u_\alpha$, $1/u_\alpha$). In addition, there are $M$ anomalous
roots each having the same multiplicity $|W|$, where $W$ is the winding number of
$\phi(t)$ (we assume that it is integer). They are located at either $-g_i/(1-g_i)$ or
at $-(1-g_i)/g_i$, depending on whether $W$ is positive or negative. Here $g_i$ are
the channel transparencies given by the eigenvalues of  $z_0^\dagger P_{3} z_0$.

We can easily rederive this result within our formalism with a procedure analogous
to that in Ref.~\onlinecite{2007-VanevicNazarovBelzig}. For simplicity we consider 
only one channel with transparency $g$. Then 
\be
	n_{F}^z = (1-g)n_{F}+gn_{F}^{\phi},
\la{nFbias}
\ee
where $n_{F}^{\phi}  =e^{-i\phi}n_{F}e^{i\phi}$. The symmetry of the spectrum of
$n_F^z$ (at zero temperature) can be demonstrated with the use of the algebra of 
four operators, 
\bea
Q_+=n_F+n_F^\phi -1 \, , & & \quad Q_- =n_F-n_F^\phi\, , \\
Q_3=i [n_F,n_F^\phi]=i Q_- Q_+ \, , & & \quad C=Q_-^2=1-Q_+^2\, .
\nonumber
\eea
The operators $Q_+$, $Q_-$, and $Q_3$ anticommute with each other, and the operator
$C$ commutes with them (this algebra relies only on $n_F$ and $n_F^\phi$ being projectors). 
Obviously $C$ also commutes with 
\be
n_F^z = \frac{1}{2}+\frac{1}{2} Q_+ + \left(\frac{1}{2}-g \right) Q_- =
\frac{1}{2}\pm\sqrt{\frac{1}{4}-g(1-g) C }
\label{n-C-square-root}
\ee
and they can be diagonalized simultaneously [the last equality in (\ref{n-C-square-root})
should be understood as a relation between eigenvalues, with different sign choices for
different eigenvectors].
Let $\Psi$ be a common eigenvector of $C$ and $n_F^z$
with the eigenvalues $C_\alpha$ and $n_\alpha^z$, respectively. 
The vector $Q_3\Psi$, if nonzero, has
the eigenvalue $1-n_\alpha^z$, which produces [according to (\ref{uz})] 
a pair of ordinary roots ($u_\alpha$, $1/u_\alpha$).
If $Q_3\Psi= 0$, then one easily proves that either $Q_-\Psi$ or $Q_+\Psi$ is zero. If
$Q_-\Psi=0$, then $C_\alpha=0$, and this corresponds to  $n_\alpha^z$ equal 0 or 1, without any
contribution to FCS. Finally, the zero modes $Q_+\Psi=0$ produce an anomalous contribution
with $C_\alpha=1$.

One can prove that, for a phase winding $W$, there are exactly $|W|$ zero
modes of $Q_+$. Assuming a non-negative $W$ (without loss of generality),
zero modes of $Q_+$ must be simultaneously eigenstates of $n_F$ and of
$n_F^\phi$ with eigenvalues 1 and 0, respectively. Such eigenstates are
easily constructed explicitly  by using the decomposition $e^{i\phi(t)} = 
e^{i [W \omega t + \phi_{+}(t) +\phi_{-}(t)]}$, where $\phi_{\pm}$ are 
the positive- and negative-frequency parts, and $\omega$ is the driving frequency.
One finds that the space of zero modes is then spanned by
$\psi_k = e^{-ik\omega t - i\phi_-(t)}$ for $k=1,2,\ldots, W$.
The corresponding eigenvalue $n_\alpha^z$ does not
have a pair, but is $W$-fold degenerate $n_\alpha^z=1-g$. Similarly, for
$W<0$, one finds $|W|$ zero modes of $Q_+$ with $n_F=0$ and $n_F^\phi=1$,
which produces $n_\alpha^z=g$.

In terms of the generating function, the separation into
the inversion-symmetric (ordinary) and anomalous (depending only on the
average charge transfer) parts may be written
as (for $W\ge 0$)
\be
	\chi(u) = \left[1+g(u-1)\right]^{W} \chi_\mathrm{inv}(u), \quad
	\chi_\mathrm{inv}(u^{-1}) =\chi_\mathrm{inv}(u) .
\la{zmc} 
\ee
With some algebraic manipulations, one can also derive
an explicitly inversion-symmetric formula for $\chi_\mathrm{inv}$,
\be
	\chi_\mathrm{inv}(u) = 
\det\left[1+g(1-g)(u+u^{-1}-2)(1-n_{F})n_{F}^{\phi}\right]
\ee
(this expression also assumes $W\ge 0$). 
This formula may be viewed as a compact form of the product over 
eigenvalues obtained 
in Ref.~\onlinecite{2007-VanevicNazarovBelzig}. Note that this separation
of the two contributions is specific to the ``bias-voltage'' case at zero
temperature.

\paragraph {9. Conclusion.}

We have derived constraints on the charge-transfer statistics between
two conductors in the limit of instant scattering and at zero temperature.
Our findings generalize the results obtained previously for the case
of a fixed scatterer with a time-dependent 
voltage \cite{2007-VanevicNazarovBelzig}. The allowed statistics is
characterized by negative real zeroes of the characteristic function
$\chi(u)$. While we have not performed a similar analysis of the
problem at finite temperature, we conjecture that the singularities
of $\log\chi(u)$ remain restricted to the negative real axis even
at finite temperature, but develop a cut instead of a discrete set
of branching points. The available simplest examples of the
charge-transfer statistics at finite temperature support this
conjecture.

Similarly to Ref.~\onlinecite{2007-VanevicNazarovBelzig}, our result (\ref{chi-result1}) 
can be interpreted as a decomposition of the charge transfer into independent tunneling 
events -- binomial processes. The effective channel transparencies 
of these events are the eigenvalues of $n_{F}^{z}$ depending on the 
time evolution of the scattering matrix. This interpretation suggests 
that the physical origin of this result lies in the assumed absence 
of interactions between electrons and that it may break down once interaction
is taken into account \cite{Lesovik-thanks}.

\paragraph{Acknowledgment.} We have  benefited from discussions
with S.~Bieri, G.~Lesovik, L.~Levitov, Yu.~Makhlin, and A.~Shytov.  
A.G.A. is grateful to ITP, EPFL for hospitality in Spring 2007.  
The work of A.G.A. was supported by the NSF under the grant DMR-0348358.




\begin{thebibliography}{15}
\bibitem{discrete-spectrum}
For any eigenstate $\Psi$ of $n_F^z$, one has $\langle\Psi | n_F^z | \Psi \rangle
= \left| n_F z | \Psi \rangle \right|^2>0$, and similarly   
$\langle\Psi | (1 - n_F^z) | \Psi \rangle > 0$. The discreteness of the spectrum follows
from the convergence of $\Tr n_F^z (1 - n_F^z) $ which can be shown under the same condition
as in~\cite{smooth}.
\end{thebibliography}
\end{document}